\documentclass[12pt, letterpaper, ]{article}

\usepackage[utf8]{inputenc}
\usepackage{amssymb}
\usepackage{multicol}
\usepackage[margin=2.4cm]{geometry}
\usepackage{amsmath}
\usepackage{amsfonts}
\usepackage{enumitem}
\usepackage{amssymb}
\usepackage{makeidx}
\usepackage{graphicx}
\usepackage[unicode=true,breaklinks=true,hidelinks]{hyperref}
\usepackage{csquotes}
\usepackage{setspace}
\usepackage{graphicx}
\usepackage{url}
\usepackage{multicol}
\usepackage{color}
\usepackage{hyperref}
\graphicspath{ {images/} }
\usepackage{tikz}
\usepackage{amsmath}
\usepackage{centernot}
\usepackage{float}
\newcommand{\xmark}{\text{\ding{55}}}
\usepackage{pifont}

\graphicspath{ {images/} }
\setlist[enumerate]{label*=\arabic*.}
\makeindex
\date{\bigskip(03.03.2023)}
\author{}
\begin{document}

\makeatletter
\def\@fnsymbol#1{\ensuremath{\ifcase#1\or *\or 
   \mathsection\or \mathparagraph\or \|\or **\or \dagger\dagger
   \or \ddagger\ddagger \else\@ctrerr\fi}}
\makeatother

\title{Bivalent Quantum Indeterminacy}

\author{Claudio Calosi\thanks{Department of Philosophy, University of Geneva} \and Iulian D. Toader\thanks{Institute Vienna Circle, University of Vienna }}

\maketitle

\begin{abstract}

\bigskip \bigskip

This paper provides a novel metametaphysical approach to quantum indeterminacy. More specifically, it argues that bivalent quantum logic can successfully account for this kind of indeterminacy, given the non-truth-functional character of its disjunction. Furthermore, it suggests that the determinable-based account of quantum indeterminacy illustrates precisely this possibility.

\bigskip \bigskip


\end{abstract}

\bigskip \bigskip

\tableofcontents

\bigskip \bigskip


\section{Introduction}

Quantum Mechanics (QM) seems to offer examples of at least two kinds of \textit{indeterminacy}. One is \textit{identity} indeterminacy, as displayed in the alleged lack of identity criteria for quantum particles.\footnote{See Lowe (1994), French \& Krause (2006) and references therein.} The other is \textit{observable} indeterminacy, illustrated by the failure of value definiteness for quantum observables. These two kinds of indeterminacy are interestingly related, but logically independent.\footnote{For the logical independence of these two kinds of indeterminacy, see Darby (2014).} We will only focus on the latter here (henceforth referred to by ``QI''). It is typically accepted that standard QM violates the classical supposition that all observables of a quantum system have definite values at all times. Furthermore, it is usually thought that, if there is QI, then this is \textit{metaphysical} in nature.\footnote{We will \textit{assume} this here, although we realize that this is not entirely uncontroversial. For skepticism about the very existence of QI see Glick (2017, 2022).} In other words, the very source of QI is the world itself. This is neither due to our knowledge of the world, nor to any semantic features of the language we use to describe the world. 

There exist currently several independent accounts of metaphysical QI in the literature,\footnote{See, for instance, Calosi and Wilson (2019, 2021, 2022), Darby and Pickup (2020), Fletcher and Taylor (2021), Torza (2020, 2021), to mention merely some that are among the most influential ones.} and there are several ways in which these accounts can be classified. In this paper, we provide a novel metametaphysical approach to this landscape. This is important not only on its own, but also for its wide-ranging implications. In particular, we will argue, our approach suggests that QI should not be understood to require, in general, a \textit{gap in logical space}, let alone a \textit{truth-value} gap.\footnote{The widely used expression ``a gap in logical space'' might be, of course, metaphorical. For one may surely maintain that, strictly speaking, there is no such thing as a logical space. We will go over some details in the next section, where we also relate this expression to the semantics of a logical calculus and the set of truth-values assigned to its sentences.} This implication will turn out to favor some accounts of QI over others -- the Aristotelian accounts -- which stipulate that in order to account for QI, the principle of bivalence (henceforth, \textsc{Bivalence}) must be dropped.\footnote{Aristotelian accounts of QI are, of course, modeled upon Aristotle's own approach to the indeterminateness of future contingents. See, e.g., Dalla Chiara, Giuntini, and Greechie (2004, 38).} To support our approach, we assume the framework of quantum logic with a standard, orthomodular lattice semantics. This will further allow us to show that beside quantum logical accounts of QI that (i) are meta-level and (ii) drop \textsc{Bivalence}, alternative accounts exist that (i) are object-level and (ii) retain \textsc{Bivalence}. 

While a detailed comparison of these alternatives is deferred to a companion paper, a crucial point in our argument in the present paper will be that the claim according to which the value indefiniteness of quantum observables requires the rejection of \textsc{Bivalence} is based on a conflation of \textsc{Bivalence} and the truth-functionality of logical connectives (in particular, the truth-functionality of quantum disjunction). Once this conflation is cleared, we can see that a non-truth-functional disjunction, i.e., one that can be true even though all its disjuncts are false, is enough to model QI in QM without introducing any truth-value gap.

\section{Logic and Placement}

There are (at least) two different axes relative to which the debate on QI has been mapped. One such axis is the \textit{Logical Axis}, as we shall call it. Here different accounts are presented and contrasted by looking at the underlying logic they use to model QI. The main alternatives are \textit{classical logic} ($\mathcal{CL}$) and \textit{quantum logic} ($\mathcal{QL}$):\footnote{See Torza (2022). This is not to say that one might not attempt to model QI with, say, intuitionistic logic in the background, e.g., as articulated in Landsman (2017).}

\begin{description}
\item \textsc{Classical Logic}: The underlying logic of QI is (still) classical logic.\footnote{Classical logic has been endorsed in Barnes and Williams (2011), Bokulich (2014), Calosi and Wilson (2019), and Schroeren (2021). Naturally, it might be not full-blown classical logic. For example, it is a  substantive question whether any form of supervaluationism can retain disjunctive syllogism.}
\item \textsc{Quantum Logic}: The underlying logic of QI is quantum logic.\footnote{Quantum logic has been discussed in the context of the debate on QI, most recently, in Calosi and Wilson (2021), and explicitly endorsed in Torza (2020, 2021) and Fletcher and Taylor (2021).}
\end{description}

Another axis is the \textit{Placement Axis}, as we shall call it. This is reminiscent of a distinction according to which there are two different strategies to accommodate metaphysical indeterminacy in general.\footnote{For the distinction between these strategies, see Wilson (2013).} In particular, different accounts differ as to whether they \textit{place} metaphysical indeterminacy at the object-level or at the meta-level: 

\begin{description}

\item \textsc{Object-Level Accounts}: The world is indeterminate, in that it contains (determinately) indeterminate states of affairs.\footnote{In the particular quantum case, Calosi and Mariani (2020) classifies different accounts in terms of their placement of QI, that is, along the \textit{Placement Axis}.  Meta-level accounts include \textit{metaphysical supervaluationism} (Barnes and Williams, 2011) and the accounts in Darby and Pickup (2020), Torza (2020, 2021), and Flecther and Taylor (2021). Arguably, the most detailed object-level account is the so-called \textit{determinable-based} account (Wilson, 2013), developed in Bokulich (2014), Calosi and Wilson (2020, 2021), Calosi and Mariani (2021), Mariani (2021) and Schoreren (2021). Note that we use ``facts'' and ``states of affairs'' interchangeably here.}
\item \textsc{Meta-Level Accounts}: The world is determinate, in that it does not contain any indeterminate state of affairs. Indeterminacy results in (i) either being indeterminate which determinate state of affairs obtains, or (ii) the world being such that for a given determinate state of affairs neither it nor its negation obtains.\footnote{Options (i) and (ii) correspond to paradigmatic meta-level accounts, namely \textit{metaphysical supervaluationism} and the \textit{gap in logical space} account, as we shall call it. More on this shortly. Note that we do not want to suggest that (i) and (ii) exhaust all possible meta-level accounts. We consider these two options because they are arguably the most widely held and discussed in the literature.}
\end{description}

Absent any substantive arguments, one may think that all combinations are possible, as per Table 1:

\begin{table}[H] \small

\centering \singlespacing
 \begin{tabular}{c|c | c}
	& \textsc{Classical Logic}   & \textsc{Quantum Logic}   \\
	& &\\
	\hline 
 	& &\\
	\textsc{Meta-Level} & $\checkmark$ & $\checkmark$ \\

	& & \\
	\hline 
	& & \\
	\textsc{Object-Level} & \checkmark &  \checkmark \\

\end{tabular}
\bigskip
\caption{Classification of Accounts of QI---First Stab}
\end{table}

However, this classification can be further refined by taking into account the following claim: if indeterminacy is to be construed as a \textit{gap in logical space}---as per (ii) in \textsc{Meta-Level Accounts} above---then the underlying logic cannot be classical.\footnote{The argument for this claim is due to Torza (2021). We also follow Torza (2022) in the characterization of the relevant notions.} We will mostly focus on this very particular way of cashing out indeterminacy, that is, as a gap in logical space. In light of this, a few clarifications are in order. 

A logical space is, roughly, a space of possibilities. It can be represented by a structure $\mathcal{S} = \langle S, @, T_S, \circ_i S \rangle$ where $S$ is a set of worlds, $@ \in S$ is the actual world, facts $F_i$ are sets of states that obtain at world $w$ iff $T_S(F,w)$ holds, and $\circ_i$ are operations on facts $F_i$. A gap in logical space is then defined as follows:

\begin{description}
\item \textsc{Logical Gap}:  There is a \textit{gap in logical space} 
iff there is a fact $F$ such that neither $F$ nor its negation $\neg F$ obtains.\footnote{See Torza (2021: 9812). \textsc{Logical Gap} is further refined such that the relevant negation operation is considered \textit{natural}, i.e., applicable to facts (Torza 2021: 9823). We shall leave this complication aside. But see footnote 5 above.}
\end{description}

\noindent This suggests that one can characterize \textsc{Indeterminacy} in terms of \textsc{Logical Gap} directly:

\begin{description}
\item \textsc{Indeterminacy$_{LG}$}: Indeterminacy occurs whenever there is a gap in logical space, 
as defined in \textsc{Logical Gap}.

\end{description}

Note that, according to \textsc{Indeterminacy$_{LG}$} no indeterminate state of affairs ever obtains. Hence, \textit{the logical gap} account is indeed a meta-level account. This is not equivalent to the other meta-level account mentioned above (in footnote 13), \textit{metaphysical supervaluationism}---which is (i) in \textsc{Meta-Level Accounts}. According to the latter:

\begin{quote} \singlespacing
 It is metaphysically indeterminate whether \textit{P} \textit{iff} there are two possibly admissible, exhaustive and exclusive state of affairs, the state of affairs that $p$ and the state of affairs that $\neg p$, and it is \textit{indeterminate} which one obtains (Barnes \& Williams, 2011: 113-114).
\end{quote}

\noindent Interestingly, whenever there is indeterminacy according to \textsc{Indeterminacy$_{LG}$}, there is indeterminacy according to \textit{metaphysical supervaluationism}. But the converse does not hold.\footnote{See Torza (2022) for an argument to this effect.} We will mostly focus on \textsc{Indeterminacy$_{LG}$} here because it is arguably the best variant of a meta-level account. Indeed, it is widely acknowledged that \textit{metaphysical supervaluationism}, at least in its original formulation, is unable to account for QI.\footnote{See among others Darby (2010), and Skow (2010). For recent proposals to amend metaphysical supervaluationism so as to face the quantum threats, see Darby and Pickup (2021)---with a reply in Corti (2021)---and Mariani, Michels \& Torrengo (2021).} If we then restrict our attention to \textsc{Indeterminacy$_{LG}$} we obtain the following (Table 2):

\begin{table}[H] \small

\centering \singlespacing
 \begin{tabular}{c|c | c}
	& \textsc{Classical Logic}   & \textsc{Quantum Logic}   \\
	& &\\
	\hline 
 	& &\\
	\textsc{Meta-Level Gap} & $\xmark$ & $\checkmark$ \\
	& & \\
	\hline 
	& & \\
	\textsc{Object-Level} & \checkmark &  \checkmark \\
	
\end{tabular}
\bigskip
\caption{Classification of Accounts of QI---Second Stab}
\end{table}

Here we will be interested solely in quantum logic, so we will leave the classical logic camp aside for now. As is well known, $\mathcal{QL}$ admits of different truth-valuational semantics. In particular, it can be associated with a \textsc{Bivalent} or a  \textsc{Non-Bivalent} semantics. Once again, absent any arguments all possible combinations should be initially considered viable, as per Table 3:

\begin{table}[H] \small

\centering \singlespacing
 \begin{tabular}{c|c | c}
	& \textsc{Bivalent} $\mathcal{QL}$ & \textsc{Non-Bivalent} $\mathcal{QL}$ \\
	& &\\
	\hline 
 	& &\\
		\textsc{Meta-Level Gap} & $\checkmark$ & $\checkmark$ \\
	& & \\
	\hline 
	& & \\
	\textsc{Object-Level} & \checkmark &  \checkmark \\
	
\end{tabular}
\bigskip
\caption{Quantum Logics for QI}
\end{table}

\noindent However, one possible way to understand the notion of a gap in logical space, perhaps \textit{the} most straightforward way given the formulation of \textsc{Logical Gap}, is as a \textit{truth-value gap}. This is sometimes explicitly acknowledged:

\begin{quote} \singlespacing
[E]very case in which it is indeterminate whether a quantum system has property
$F$ manifests itself in a \textit{truth-value gap} with respect to the proposition ascribing $F$
to that system (Fletcher and Taylor, 2021a: \S1, italics added).
\end{quote}

\noindent Indeed, one could provide a semantic characterization of indeterminacy using exactly the notion of a truth-value gap:

\begin{description}
\item \textsc{Indeterminacy$_{TVG}$}: Indeterminacy occurs whenever there is a truth-value gap, i.e., a sentence $p$ whose truth-value is indeterminate, that is, neither true nor false.
\end{description}

\noindent In fact, under certain assumptions, \textsc{Indeterminacy$_{LG}$} and \textsc{Indeterminacy$_{TVG}$} are logically equivalent.\footnote{See Torza (2022) for an argument supporting this claim. The proof of the equivalence rests on two premises: (i) that the relevant object language contains no irreferential terms, and (ii) every fact is expressible in the relevant language. Note that Fletcher and Taylor (2021)---as we saw---endorse \textsc{Indeterminacy$_{TVG}$}. There is no indication that they reject either (i) or (ii) in Torza's proof of the equivalence between  \textsc{Indeterminacy$_{TVG}$} and \textsc{Indeterminacy$_{LG}$}. Hence, they arguably endorse the latter as well. This is enough to claim that theirs can be thought of as a meta-level account---which is what counts for the purpose of this paper. Naturally, this does not mean that their account of QI is equivalent to e.g., Torza's. In effect it is not. They add that ``for any such case of quantum indeterminacy it must be \textit{possible} for the aforementioned proposition [the proposition that ascribes a particular property to a given system in a case of indeterminacy, which is indeterminate in truth value being neither true nor false] to be true'' (Flecther and Taylor (2021: \S1). This makes for a substantive difference between the accounts. See Fletcher and Taylor (2021: \S4.4).} In this case, the obvious choice for the proponent of a meta-level account of indeterminacy is the Aristotelian one: drop \textsc{Bivalence}. This leads one to turn to \textsc{Non-Bivalent} $\mathcal{QL}$ as a framework for QI, which would reshape the landscape as in Table 4:

\begin{table}[H] \small

\centering \singlespacing

 \begin{tabular}{c|c | c}
	& \textsc{Bivalent} $\mathcal{QL}$ & \textsc{Non-Bivalent} $\mathcal{QL}$ \\
	& &\\
	\hline 
 	& &\\
    	\textsc{Meta-Level Gap} & $\xmark$ & $\checkmark$ \\
	& & \\
	\hline 
	& & \\
	\textsc{Object-Level} & \checkmark &  \checkmark \\
	
\end{tabular}
\bigskip
\caption{Restricting Quantum Logics for QI}
\end{table}

In what follows we provide novel arguments for the following claim: dropping \textsc{Bivalence} is \textit{not necessary} to account for QI. Against Aristotelian accounts, we submit that the non-truth-functionality of quantum disjunction is capable, \textit{by itself}, to express indeterminate states of affairs, and thus to do full justice to QI. Indeed, we also suggest that \textsc{Non-Bivalent} $\mathcal{QL}$ makes for a worse option overall. This is not important just on its own. That very argument also points towards a more general---and arguably deeper---conclusion, namely that QI should \textit{not be seen as a gap in logical space after all}. Thus, we believe this constitutes an indirect argument in favor of object-level accounts. Indeed, as we pointed out, \textsc{Meta-Level Gap} arguably represents the best version of a meta-level account of QI. Moreover, we will see how the paradigmatic example of an object-level account, the \textit{determinable-based} account, can benefit from the argument in this paper.\footnote{We discuss \textsc{Object-Level} accounts of QI in the framework of \textsc{Non-Bivalent} $\mathcal{QL}$ in a companion paper. Such an account, in Stairs (1983, 2016), introduces indeterminate states of affairs as primitive disjunctive facts, that is, facts described by a non-truth-functional quantum disjunction, i.e., a disjunction that can be true despite not having any true disjuncts. We agree that primitive disjunctive facts can be described by a non-truth-functional quantum disjunction, but we take the latter to be a disjunction that can be true despite having all and only false disjuncts. Thus, on our view, as will be clear, \textsc{Bivalence} and the value indefiniteness of observables can cohabit peacefully within the same quantum logical framework, as the \textit{determinable-based} account illustrates.} The result is that this is arguably the front-runner as an account of QI (see Table 5):

\begin{table}[H] \small

\centering \singlespacing
 \begin{tabular}{c|c | c}
	& \textsc{Bivalent} $\mathcal{QL}$ & \textsc{Non-Bivalent} $\mathcal{QL}$ \\
	& &\\
	\hline 
 	& &\\
    	\textsc{Meta-Level Gap} & $\xmark$ & $\xmark$ \\
	& & \\
	\hline 
	& & \\
	\textsc{Object-Level} & \checkmark & \xmark \\
	
\end{tabular}
\bigskip
\caption{The Front-Runner Account of QI}
\end{table}

Before we reach this result, we need to introduce---albeit briefly---$\mathcal{QL}$. 


\section{Quantum Logic}

We present here the quantum logical notions that are needed in order to run the main argument in the rest of the paper. This is obviously not meant as a substitute for a comprehensive introduction to quantum logic.\footnote{ For a classic such introduction, see e.g. Beltrametti and Cassinelli (1981). An extended, although now evidently outdated, bibliographical list, is Pavi\v{c}i\'{c} (1992). For a more recent extended survey, see Dalla Chiara, Giuntini, and Greechie (2004). For an abstract presentation of quantum logic, in the algebraic setting of Dunn and Hardegree (2001), see Horvat and Toader (2023).} But we include this so that the paper becomes as self-contained as possible. Even so, we limit ourselves to taking orthomodular lattices as the standard algebraic semantics of quantum logic, thus basically ignoring Kochen and Specker's partial Boolean algebras as an alternative semantics.\footnote{Although this will be inconsequential here, we consider $\mathcal{QL}$ as a natural deduction system, rather than as an axiomatic system in the Hilbert-Ackermann style or as a sequent calculus in the Gentzen style.}

Let $\mathcal{QL}$ be a language with variables for sentences, $p$, $q$, ..., and symbols for connectives, $\neg, \wedge, \vee $, such that the set of sentences is defined by induction: for any sentences $p$ and $q$ in the set, $\neg p$, $p \wedge q$, and $p \vee q$ are also in that set.\footnote{Of course, one can also introduce a symbol, $\rightarrow$, for a conditional connective that can be defined as follows: $ p \rightarrow q $ iff $ \neg p \vee ( p \wedge q ) $. This is the so-called Sasaki hook, which is a counterfactual conditional, and is weaker than the classical one, since for example it does not admit strengthening the antecedent. But for our purposes here, introducing the conditional is entirely optional.} As we understand it here, $\mathcal{QL}$ has all the rules of a classical natural deduction system, like double negation, introduction and elimination rules, modus ponens and modus tollens, and De Morgan rules. However, unlike a classical system, $\mathcal{QL}$ restricts, for example, $\vee$-elimination such that one cannot use side premises when applying the rule.\footnote{In other words, $\vee$-elimination  takes the form ``if $p$ implies \textit{r} and $q$ implies \textit{r}, then $p\vee q$ implies \textit{r}'', rather than ``if $\{p, s\}$ implies \textit{r} and $\{q, t\}$ implies \textit{r}, then $\{s, t, p \vee q\}$ implies \textit{r}''. This restriction is equivalent to the weakening of the classical conditional noted in the previous footnote (Humberstone 2011, 300), but it preserves all theorems when disjunction is the only connective in the language (\textit{op. cit}, 918).}

Further, let $\mathcal{LA}$ be an ortholattice, i.e., a partially ordered set of elements, together with operations $^{\bot}, \cap, \cup$, for orthocomplementation, meet, and join, respectively. Meet is just set-theoretical intersection. Join is not set-theoretical union, but the operation that yields the smallest element containing the joined elements, i.e., their span. Orthocomplementation is defined in terms of a map $h : \mathcal{QL} \longrightarrow \mathcal{LA}$ as follows:

\begin{center}

  $h(p) = h(q)^{\bot}$ iff $\{x : x \subseteq h(p)\} = \{x : x {\bot} h(q) \}$
  
\end{center}

We assume that \textit{h} is a homomorphism, which allows $\mathcal{LA}$ to be taken as a model of $\mathcal{QL}$:

\begin{center}
    
$h(\neg p) = h(p)^{\bot}$ \\ 
$h(p \wedge q) = h(p) \cap h(q)$ \\ 
$h(p \vee q) = h(p) \cup h(q) $ 

\end{center}

This also yields a definition of logical consequence:

\begin{center}
  
  $\Gamma \models q$ iff $\cap \{h(p):p \in \Gamma \} \subseteq h(q)$.

\end{center}

We should note that $\mathcal{LA}$ validates the rule of orthomodularity: 

\begin{center}
    if $h(p) \subseteq h(q)$, then $h(q) = h(p) \cup (h(p)^{\bot} \cap h(q))$.\footnote{This is a version of Dedekind's rule of modularity, which was originally introduced for philosophical and mathematical reasons to replace the rule of distributivity (Birkhoff and von Neumann 1936, 832), but was subsequently replaced by the weaker rule of orthomodularity. It is easy to see that the latter is equivalent to modus ponens, if the conditional is the Sasaki hook (Humberstone 2011, 302).}
\end{center}

As is well known, it turned out that the structure of our orthomodular lattice $\mathcal{LA}$ is precisely that of the Hilbert lattice, i.e., the structure of the closed linear subspaces of an infinite-dimensional Hilbert space. This may be taken to justify taking $\mathcal{QL}$ as a logical calculus for standard quantum mechanics. It also turned out that some connectives in $\mathcal{QL}$ are not truth-functional, so that, e.g., ortho-disjunctive statements may be true even if all their ortho-disjuncts are false.\footnote{Since we take $\mathcal{LA}$ as a model of $\mathcal{QL}$, we follow Lloyd Humberstone (\textit{op. cit.}) in calling quantum disjunction an ``ortho-disjunction''.} This can be shown as follows.

Let \textit{g} be a truth-valuation, $g : \mathcal{LA} \longrightarrow \{T,F\}$. Provably, \textit{g} is not a homomorphism.\footnote{See Hellmann 1980 for a proof. The fact that $g$ is not a homomorphism also follows immediately from the Kochen-Specker theorem. More exactly, in Kochen and Specker's algebraic framework, where connectives are defined only for a subset of all elements of the orthomodular lattice, the theorem guarantees that the partial Boolean algebra of the Hilbert space has no homomorphic Boolean extension, i.e., there exists no homomorphism from that algebra to the two-valued Boolean algebra $\{T,F\}$, on the assumption that the space is of dimension greater than two (Kochen and Specker 1967).} Considering another truth-valuation $f : \mathcal{QL} \longrightarrow \{T,F\}$, where $f = g \circ h$ such that $f(p) = g(h(p))$, it follows that \textit{f} is not a homomorphism, either. There are sentences $p, l, n, k \in \mathcal{QL}$ such that $f(p) = f(l) = f(n) = f(k) = F$, but $f(p \vee l) = T$ and $f(n \vee k) = F$. What this means is that $\mathcal{QL}$ is not a categorical logical calculus with respect to a Boolean semantics, as it admits distinct, non-isomorphic truth tables for ortho-disjunction.\footnote{In this respect, ortho-disjunction is closer to classical disjunction than typically appreciated, if the latter is taken under Carnap's non-normal interpretation of $\mathcal{CL}$ (see Carnap 1943. For a rigoros comparative analysis of classical and quantum connectives, see Horvat and Toader 2023.) In a companion paper, we raise this point against the view (defended, e.g., in Bigaj 2001) that an account of QI requires dropping \textsc{Bivalence} because a disjunction that can be true despite not having any
true disjuncts is more classical than a disjunction that can be true despite having only false disjuncts.}

One can easily see that the fact that ortho-disjunction is not truth-functional entails, in the other direction, that the rule of distributivity is false in $\mathcal{QL}$. Suppose we perform a $z$-spin measurement on a physical system (say, a spin-one particle) in a $z$-spin up eigenstate. Let sentence $s_{z}^{\uparrow} \in \mathcal{QL}$ state the result. Then we get the following derivation:

\bigskip

1. $ s_{z}^{\uparrow} \wedge (s_{x}^{\uparrow} \vee s_{x}^{\downarrow}) $

\smallskip

2. $ \neg(s_{z}^{\uparrow} \wedge s_{x}^{\uparrow}) \wedge \neg(s_{z}^{\uparrow} \wedge s_{x}^{\downarrow}) $

\smallskip

3. $ \neg((s_{z}^{\uparrow} \wedge s_{x}^{\uparrow}) \vee (s_{z}^{\uparrow} \wedge s_{x}^{\downarrow})) $

\bigskip

Assuming that the Uncertainty Principle justifies premise 2, this immediately give us a counterexample to distributivity, which can be stated as follows:

\begin{center}
    
$ \ulcorner s_{z}^{\uparrow} \wedge (s_{x}^{\uparrow} \vee s_{x}^{\downarrow}) \urcorner \nvdash \ulcorner (s_{z}^{\uparrow} \wedge s_{x}^{\uparrow}) \vee (s_{z}^{\uparrow} \wedge s_{x}^{\downarrow}) \urcorner $

\end{center}

Clearly, if $ \ulcorner s_{x}^{\uparrow} \vee s_{x}^{\downarrow}\urcorner$ is true even though $s_{x}^{\uparrow}$ and $s_{x}^{\downarrow}$ are both false at a $z$-spin up eigenstate, then the counterexample to distributivity is true.\footnote{As we shall see, this is exactly the analysis that, we believe, the proponent of a particular account of QI -- the \textit{determinable-based} account -- can and should provide.} This is validated by our semantics, $\mathcal{LA}$:

\begin{center}
    
    $h(s_{z}^{\uparrow}) \cap (h(s_{x}^{\uparrow}) \cup h(s_{x}^{\downarrow}))  \nsubseteq h(s_{z}^{\uparrow} \cap s_{x}^{\uparrow}) \cup h(s_{z}^{\uparrow} \cap s_{x}^{\downarrow})$

\smallskip

    $h(s_{z}^{\uparrow}) \cap \textbf{1} \nsubseteq \varnothing \cup  \varnothing $

\smallskip

    $h(s_{z}^{\uparrow}) \nsubseteq \varnothing $
    
\end{center}

Now we are ready to tackle our main question. If the world has QI, and $\mathcal{QL}$ is taken as an appropriate logical framework for modeling QI, what are the implications for the metaphysical accounts of QI classified in the previous section? In particular, we are concerned to determine whether these accounts require a ``gap in the logical space'' or, even more specifically, a ``truth-value'' gap which would force one to drop \textsc{Bivalence}.

\section{Bivalence or Non-Bivalence?}

We want to argue here that taking ortho-disjunction to be non-truth-functional is enough to account for QI on its own. Hence, against Aristotelian accounts, there \textit{is no need} to reject \textsc{Bivalence}---at least not one coming from QI or QM. One may, of course, do so, that is one may consider a rejection of \textsc{Bivalence} as sufficient for providing a metaphysical account of QI. But that is by no means necessary. Furthermore, we also want to show that if the rejection of \textsc{Bivalence} is understood to be a consequence of QM, then it is rather ill motivated. However, its rejection is just as poorly justified, if it is taken to be entailed by the very structure of the orthomodular lattice $\mathcal{LA}$, defined above. Finally, we also briefly point out that the rejection of  \textsc{Bivalence} is optional even where it was most often regarded as indispensable, i.e., in dissolving the sorites paradox.

It has sometimes been suggested that \textsc{Bivalence} must be given up as a consequence of the Kochen-Specker theorem.\footnote{One can find such a view, for instance in Bell and Hallett (1982). Others draw the same conclusion from the Jauch-Piron theorem (1963): ``Jauch and Piron show
that any so-called orthomodular lattice (in particular any Hilbert lattice) admits
total homomorphisms onto $\{0, 1\}$ iff it is distributive. Note that this means that any form of quantum logic \textit{must give up bivalence} (Bacciagaluppi 2009: 56). However, drawing such a conclusion indicates what we will shortly call \textsc{Conflation}.} However, what the Kochen-Specker theorem implies is not the falsity of \textsc{Bivalence}, but the failure of distributivity (and thus of truth-functionality). William Demopoulos was the first to point this out, repeatedly, starting with his PhD thesis: 

\begin{quote} \singlespacing
The fact that there are no homomorphisms $g : \mathcal{LA} \longrightarrow \{T,F\}$ must
be \textit{sharply distinguished} from the question of the bivalence
of the language \textit{L} in which the theoretical propositions, are formulated (Demopoulos 1974: 33---modified for unified notation, italics added).
\end{quote}

The Kochen-Specker theorem implies, to be sure, that the ortho-connectives are not truth-functional, and that na\"{i}ve (or classical) realism, i.e., a realism that requires a Boolean structure of physical properties at any given time, is false. But neither of these two implications entails the rejection of \textsc{Bivalence}. Demopoulos summarizes the point as follows: 

\begin{quote}
\singlespacing
There are two different accounts of indeterminism which are historically important.\footnote{Clearly, Demopolous uses ``indeterminism'' for what we mean by ``indeterminacy''.} The first, which apparently goes back to Aristotle, rejects bivalence: A theory is indeterministic if it assumes that there are propositions whose truth value is indeterminate. The second, represented by the quantum theory,
retains bivalence while rejecting semi-simplicity. An indeterministic theory is then characterized by the absence of two-valued homomorphisms, and therefore, of two-valued measures. The coherence of indeterminateness seems to rest on the Aristotelian metaphysic of act and potency. But nothing of this sort is required by the indeterminism of quantum mechanics. This form of indeterminism implies that there is no
Boolean representation of the properties obtaining at a given time; yet for any property \textit{P} it is completely determinate whether or not \textit{P} holds. (Demopoulos 1976, 76sq). 
\end{quote}

Demopoulos' ``semi-simplicity'' denotes an algebraic property equivalent to the existence of  homomorphisms $g : \mathcal{LA} \longrightarrow \{T,F\}$, i.e., precisely the property that the Kochen-Specker theorem essentially denies $\mathcal{LA}$ can have. Thus, as Demopoulos argued, if one construed \textsc{Bivalence} in terms of semi-simplicity, then of course the theorem would entail the rejection of \textsc{Bivalence}. But, so his argument goes, there is no reason QM should commit one to such a construal of \textsc{Bivalence}. Taking this into account then, the point of this paper may be expressed as follows: 

\begin{description} \singlespacing
\item \textsc{Conflation}: Metaphysical accounts of QI that drop \textsc{Bivalence} conflate the two senses of indeterminacy---namely, failure of \textsc{Bivalence} and failure of truth-functionality.
\end{description}

However, our considered view is that QI need not, and perhaps should not, be regarded as Aristotelian indeterminateness. It is one thing to say that it is true that ``This photon will decay tomorrow'' or ``This photon will not decay tomorrow'', while each of these statements is neither true nor false, and another thing to say that it is true that ``This photon passed through the upper slit'' or ``This photon passed through the lower slit'', when each of these statements is false.\footnote{Naturally, indeterminacy of future contingent events need not require dropping \textsc{Bivalence} either. For some recent examples see, among others, Cameron (2015), Correia and Rosenkranz (2018), and Grandjean (2021)---and references therein.}

Nevertheless, the Aristotelian attack on \textsc{Bivalence} has recently resurfaced. The claim this time is that the very structure of the closed subspaces of the Hilbert space, i.e., our lattice $\mathcal{LA}$, implies the falsity of  \textsc{Bivalence}.\footnote{See Fletcher and Taylor (2021) and our critical discussion thereof in the next section.} Whoever insists on preserving  \textsc{Bivalence} fails allegedly to read this structure correctly. So if one adopted $\mathcal{QL}$ as a basis for an account of QI, then one would be forced to drop \textsc{Bivalence}. But, as we will see shortly, nothing is further from the truth. Birkhoff and von Neumann’s original calculus of experimental propositions is certainly not a \textsc{Non-Bivalent} $\mathcal{QL}$. They did not reject \textsc{Bivalence}, and they did not take it that a \textsc{Non-Bivalent} semantics can be read off from the structure of the closed linear subspaces of a Hilbert space. Their informal presentation of the calculus, in algebraic-semantic terms, stipulates that if the Hilbert space associated with a quantum system is to be ``imbued with reality'', the ``closed linear subspaces of Hilbert space [must] correspond one-many to experimental propositions, but one-one to physical qualities.'' (Birkhoff and von Neumann 1936, 825-8) As far as we can see, there is no explicit or tacit implication that the algebraic structure of physical qualities requires that \textsc{Bivalence} be given up.\footnote{For the explicit construction of a bivalent semantics for $\mathcal{QL}$, see again Horvat and Toader 2023.}

\section{Defending \textsc{Bivalent} $\mathcal{QL}$}

Fletcher and Taylor (2021) claim that it is the structure of the closed subspaces of the Hilbert space, i.e., the structure of our orthomodular lattice $\mathcal{LA}$, that implies the falsity of \textsc{Bivalence}. Adopting $\mathcal{QL}$ as a basis for an account of QI allegedly forces one to drop \textsc{Bivalence}. Here is their argument in full:

\bigskip

\begin{quote} \singlespacing
    Quantum logic’s treatment of negation in terms of ortho-complementation rather than complementation simpliciter marks a critical difference between it and classical negation. For while a set and its complement taken together always exhaust the entire space, the same is not true of a subspace and its orthogonal complement: unless $P$ is either the identity or the zero operator, $ran(P) \cup ran(P)^{\bot}$ is never identical to the entire space. Accordingly, for any such P there will always be states in neither $ran(P)$ nor $ker(P)$---that is, there will always be states for which the property ascription $p$ will be neither true nor false. Thus, we see already that \textit{quantum logic's treatment of negation entails that its semantics for quantum property ascription cannot be classical: it violates the principle of bivalence} (Fletcher and Taylor 2021: \S2.2, italics added).
\end{quote}

Now, quite apart from what we have seen above Birkhoff and von Neumann may have thought about this matter, we believe that there is no compelling reason to think that a \textsc{Non-Bivalent} semantics must be read off from the structure of the closed linear subspaces of a Hilbert space. Fletcher and Taylor's claim that \textsc{Bivalence} must be given up is based on their understanding of quantum negation. Quantum negation, represented algebraically as orthocomplementation (as we have seen above: $h(\neg p) = h(p)^{\bot}$) is taken to be different than classical negation, represented as mere complementation, in that whereas the latter is such that the union of a set and its complement exhausts the entire space, the former is such that in general (that is, unless \textit{p} is either a tautology or a contradiction) the union of a set and its orthocomplement  doesn't: $h(p) \cup h(p)^{\bot}$ is ``never identical to the entire space''.

However, as we have seen in \S3, there is no union operation defined on the elements of the Hilbert lattice $\mathcal{LA}$: a union of subspaces is \textit{not} a subspace, so there is no wonder that it is not identical to the entire Hilbert space. As we have also seen, $\cup$ is not set-theoretical union, but the join operation, which yields the span of any two subspaces, that is the smallest subspace that includes them both. But $h(p) \cup h(p)^{\bot}$ is identical to the entire Hilbert space, for the smallest subspace that includes any set and its orthocomplement is actually the entire Hilbert space. \textit{Pace} Fletcher and Taylor, the semantics of negation, properly understood in terms of the orthomodular lattice $\mathcal{LA}$, does not provide a valid reason for dropping \textsc{Bivalence}. Lest one doubts the Aristotelian character of their account of QI, Fletcher and Taylor add: ``A special case of this bivalence failure will be key to our account of indeterminacy below'' (Fletcher and Taylor, 2021: \S2.2).

In a nutshell, then, our reply here is simply that \textsc{Bivalence} fails \textit{if and only if} the operation $\cup$ between $ran(P)$ and $ran(P)^{\bot} = ker (P)$ is \textit{set-theoretic} union. However, $\cup$ is the lattice operation of \textit{join}, and its result is always a \textit{span}. But the span of $ran(P)$ and $ran(P)^{\bot} = ker (P)$ is indeed \textit{the entire space}. This, we take it, is enough to undermine the argument above.

Nevertheless, there is nothing that could prevent one to allow set-theoretical union on the elements of $\mathcal{LA}$, as long as one is clear that it is only the join operation that can represent ortho-disjunction. But a close consideration of Fletcher and Taylor's understanding of ortho-disjunction reveals further complications. They do correctly note that any ortho-disjunction of sentences is algebraically represented by the span of the corresponding subspaces of the Hilbert space. Letting \textit{span} to denote the span, as they do, their semantic principle for quantum disjunction is as follows: $h(p \vee q) = span(h(p) \cup h(q))$, where $\cup$ is now set-theoretical union. However, when they turn to describing the non-truth-functional character of ortho-disjunction, Fletcher and Taylor state that what accounts for this character is the fact that $h(p) \cup h(q)$ is ``never a subspace''. But if the union of subspaces is never a subspace, then the union operation should not be taken to have any semantic significance. Therefore, it should not be able to support the argument from the semantics of ortho-negation, against \textsc{Bivalence}, in the first place.

Thus it appears that there is nothing in standard QM or in the metaphysics of QI that requires that we drop \textsc{Bivalence}. But we believe that \textsc{Non-Bivalent} $\mathcal{QL}$ ought to live on as a basis for an object-level account of QI.\footnote{Nonetheless, see footnotes 18 and 26 above.} However, it should be clear that \textsc{Bivalent} $\mathcal{QL}$ is perfectly able, by itself, to provide an arguably better account of QI as well. Indeed, we contend that an account based on \textsc{Bivalent} $\mathcal{QL}$ shows that QI \textit{need not} be understood as a \textit{truth-value} gap, and indeed as a \textit{gap in logical space at all}. In fact, insofar as \textsc{Bivalent} $\mathcal{QL}$ should be preferred, an account of QI that does not require indeterminacy to be a gap in logical space should be preferred. And we already saw in \S2 that there are accounts according to which QI is not any such gap, namely \textit{object-levels} account. Thus, the arguments in this paper are also (indirect) arguments in favor of those accounts. In the next section, we consider several objections that the Aristotelian could raise, and has raised, against the claim that the \textit{determinable-based} approach to QI offers precisely such an account.

Nevertheless, one may still have lingering doubts with respect to \textsc{Bivalence}, and one may suspect that even if it need not be dropped, it is in fact a liability for any account of QI. Famously, Quine argued that \textsc{Bivalence} ``seals the [sorites] paradox, requiring as it does at each stage that the statement that a heap remains, or that the man is bald, be univocally true or false.'' (Quine 1981, 91). Might \textsc{Bivalence} seal the fate of QI as well? We don't think so, for QI is not of the same type as the indeterminacy of the sorites variety. Furthermore, a \textsc{Bivalent} $\mathcal{QL}$ approach to the sorites paradox shows that the non-truth-functionality of disjunction, by itself, is enough to dissolve the paradox (Rumfitt 2018).\footnote{Dummett argued, more generally, that $\mathcal{QL}$ itself is problematic, because it requires a classical disjunction in the metalanguage, in the very formulation of \textsc{Bivalence} (Dummett 1976, 280). But it is not clear that  \textsc{Bivalence} must be formulated as Dummett thought it should. Along similar lines, more recently, Rumfitt argued that $\mathcal{QL}$ cannot be rationally accepted by the classical logician, for the proof that distributivity fails in QM is either unsound or rule-circular, demanding classical logic in the metalanguage (Rumfitt 2015). For a critical discussion of this argument, see Toader 2023.}

\section{A Bivalent $\mathcal{QL}$ account of QI}

\noindent The arguments above strongly support the claim that, Aristotelian accounts notwithstanding, QI need \textit{not} be construed as a gap in logical space. Indeed the \textit{determinable-based} account is better seen as refusing this construal of QI. Torza (2021) duly recognizes this, and provides an objection. Fletcher and Taylor (2021) provides different objections to that account. Some---but not all---of them target the \textit{determinable-based} account developed on the basis of $\mathcal{CL}$. As such, they are beyond the scope of the present paper. Some---but not all---target the \textit{determinable-based} account in general, that is, independently of any background logic and thus are in the present remit. It is significant that the analysis put forward here helps respond to those objections.

It is useful to rehearse briefly the core of the determinable-based account. Roughly, according to the determinable-based account, metaphysical indeterminacy involves the obtaining of an indeterminate state of affairs (SOA)---roughly a state of affairs where a constitutive object fails to have a unique determinate of a determinable:

\begin{quote} \singlespacing
        \textsc{Determinable-Based MI}: What it is for an SOA to be MI in a given respect \textit{R} at a time \textit{t} is for the SOA to constitutively involve an object (more generally, entity) \textit{O} such that (i) \textit{O} has a determinable property \textit{P} at \textit{t}, and (ii) for some level \textit{L} of determination of \textit{P}, \textit{O} does not have a unique level-\textit{L} determinate of \textit{P} at \textit{t} (Wilson, 2013: 366).
    \end{quote}

In the quantum case, this amounts to the claim that a quantum system (in a given state) has a determinable but no unique determinate of that determinable. Restricting the attention to the case in which there are only two levels of determination---hence we only have maximally unspecific determinable and maximally specific determinates---there. are two cases. Either a quantum system has a determinable but there no determinate of that determinable (\textit{gappy} indeterminacy), or it has more than one---albeit in a relativized fashion, or to a different degree (\textit{glutty} indeterminacy).\footnote{We refer the interest reader to Calosi and Wilson (2019) for details.}

Let us then look at the objections against the determinable-based account, starting with Torza's. The argument---in a nutshell---is the following. The determinable-based account entails that, as Torza writes:

\begin{quote} \singlespacing
    $w_1$. Determinables are not analyzable in terms of their maximally specific determinates (Torza, 2021: 9827).
\end{quote}

\noindent Yet, as Torza notes, one could identify, say, the determinable of ``having spin$_z$'', expressed by sentence $s_{z} \in \mathcal{QL}$, with the span of ``having spin$_z$ up'' and ``having spin$_z$ down'', expressed by the \textit{ortho-disjunction} $s_{z}^{\uparrow} \vee s_{z}^{\downarrow}$, where disjuncts refer to maximally specific determinates.\footnote{We assume with Torza that ``having spin$_z$'' is a determinable whose maximally specific determinates are ``having spin$_z$ up'' and ``having spin$_z$ down''.} This, Torza continues, establishes:

\begin{quote} \singlespacing
    $w_2$. Some determinables are analyzable in terms of their maximally specific determinates (Torza, 2021: 9828).
\end{quote}

\noindent Clearly, $w_2$ contradicts $w_1$. The arguments in this paper can be used to provide a clear response to the objection. In particular, we contend that what the \textit{determinable-based} account entails \textit{is not} $w_1$, but rather something \textit{weaker}, namely:

\begin{description}
\item[]\textsc{No Classic Disjunction}: Determinables are not analyzible in terms of \textit{classical disjunctions} of their maximally specific determinates. 
\end{description}

Torza's objection is in particular directed to QI of the \textit{gappy} variety, in which the system has a determinable but \textit{no} determinate. In this case, it is easy to see why the \textit{determinable-based} account indeed entails \textsc{No Classic Disjunction}. If ``having spin$_z$'' were to be analyzable as ``having spin$_z$ up'' \textit{or}  ``having spin$_z$ down'', where \textit{or} is a classical disjunction, then it would indeed follow that whenever a quantum system has the determinable it has at least one of its two maximally specific determinates. That is because classical negation \textit{is truth-functional}. However, our main point in this paper is exactly that failure of distributivity and, thus, of the non-truth-functionality of ortho-disjunction are enough to account for QI---within $\mathcal{QL}$. And indeed, this is exactly what the defender of the \textit{determinable-based} account \textit{could}---perhaps \textit{should}---do in this setting. She \textit{could}---should she want to---analyze quantum determinables as \textit{ortho-disjunctions} of their maximally specific determinates. This analysis \textit{does not} threaten her account of QI exactly because she could take $ \ulcorner s_{z}^{\uparrow} \vee s_{z}^{\downarrow}\urcorner$ to be true---indicating that the relevant system has the determinable ``spin$_z$---even though $s_{z}^{\uparrow}$ and $s_{z}^{\downarrow}$ are both false---indicating that the relevant system has neither the determinate ``spin$_z$ up'' nor ``spin$_z$ down''.

Fletcher and Taylor (2021) voices a similar objection, even if within the background of classical logic. They write:

\begin{quote} \singlespacing
It’s clear that, according to CW [Calosi and Wilson, proponents of the \textit{determinable-based} account], determinables are not reducible to mere disjunctions of corresponding determinates. And yet we've just seen how, in certain cases, they must view the determinable SPIN$_x$ as at least logically equivalent to the disjunction of its determinates (Fletcher and Taylor, 2021: \S4.2.4)
\end{quote}

This is relevantly similar to Torza's argument and our reply is the same. The \textit{determinable-based} account---in its \textit{gappy} variety---is committed to \textsc{No Classic Disjunction}, and thus to reject that a determinable is logically equivalent to a \textit{classical disjunction} of its determinates. But, as it stands, it is compatible---at least at first sight---with all of the following:\footnote{Note that in what follows we are slightly abusing terminology, insofar as ortho-disjunction applies to sentences of $\mathcal{QL}$.}

\begin{itemize}
    \item A quantum determinable is \textit{analyzable in terms} of an \textit{ortho-disjunction} of its maximally specific determinates.
    
    \item A quantum determinable is \textit{reducible to} an \textit{ortho-disjunction} of its maximally specific determinates.
    
    \item A quantum determinable is \textit{logically equivalent} to an  \textit{ortho-disjunction} of its maximally specific determinates.
    
    \item A quantum determinable \textit{just is} an  \textit{ortho-disjunction} of its maximally specific determinates.
\end{itemize}

Depending on one's views about \textit{reducibility}, \textit{definability},  \textit{logical equivalence} and \textit{generalized identity} the different stances above might turn out to be not equivalent and their relations highly non-trivial.

Fletcher and Taylor (2021, \S4.2.4) provides yet another argument against the \textit{determinable-based} account. The argument is roughly as follows. Consider a quantum system, an electron $e$ in Fletcher and Taylor's example, such that $e$ is not ``spin$_x$ up''. Then, ``\textit{not} being spin$_x$ up'' is a property which corresponds to a subspace of the Hilbert space of $e$ implicitly definable from ``$e$ is spin$_x$ up''---the last bit following from the extensionality of classical logic. Now, there are only four such subspaces: the null space, the entire Hilbert space, the subspace spanned by ``spin$_x$ up'', and the subspace spanned by ``spin$_x$ down''. The problem is that none will do according to the \textit{determinable-based} account. It is trivial to see that the first three subspaces won't. This leaves the fourth subspace. But, they continue, given classical logic ``$e$ is spin$_x$ down'' is logically equivalent to ``$e$ is not spin$_x$ up''. Now consider a case in which $e$ is in $\frac{1}{\sqrt{2}} (|\uparrow\rangle _x - |\downarrow\rangle_x)$. In this case, so the argument goes, either both ``$e$ is spin$_x$ down'' and ``$e$ is not spin$_x$ down'' are false (as per \textit{gappy} indeterminacy), or both are true (as per \textit{glutty} indeterminacy). Both options violate classical logic.

At this point, one may simply notice that, as Fletcher and Taylor explicitly admit, the argument targets the combination of the \textit{determinable-based} account and classical logic. So that it is unclear whether it can be marshaled against the former in the present, quantum-logical setting. But the argument can be resisted on independent grounds. Indeed, we submit, both the classical logician and the quantum logician can resist it albeit in different ways.

As we saw, classical logic and quantum logic treat negation differently. Classical negation is \textit{set-theoretic complementation}. Thus, according to the classical logician ``\textit{not} being spin$_x$ up'' does not correspond to a subspace, it is not logically equivalent to ``being spin$_x$ down'', and it is not a property after all---if properties are associated only with subspaces. Indeed, apart from a very controversial take on the nature of properties, one might---and perhaps simply should---deny that every predicable condition corresponds to a property. The paradigmatic example is ``existence''. This is a predicable condition but it is controversial at best that it corresponds to a property.

By contrast, quantum negation is \textit{ortho-complementation}. In this case ``\textit{not} being spin$_x$ up'' is logically equivalent to ``being spin$_x$ down'' and it is a property in the light of the \textit{determinable-based} account. But as we saw, there is nothing problematic within $\mathcal{QL}$ to maintain that when $e$ is in $\frac{1}{\sqrt{2}} (|\uparrow\rangle _x - |\downarrow\rangle_x)$ both ``$e$ is spin$_x$ down'' and ``$e$ is not spin$_x$ down'' are false (as per \textit{gappy} indeterminacy). Indeed, this is exactly what we suggested in this paper.\footnote{The case of \textit{glutty} indeterminacy is different. In that case, determinable based theorists are explicit in claiming that sentences of the form ``$e$ is spin$_x$ down'' are not \textit{truth-evaluable}---see e.g., Calosi and Wilson (2021: 3301-3302). What is truth-evaluable are sentences of the form ``$e$ is spin$_x$ down \textit{relative to $x$}'', or `$e$ is spin$_x$ down \textit{to a degree $d$''}. We did not  consider \textit{glutty} indeterminacy in detail so we will leave it at that. To be fair, Fletcher and Taylor lists yet another objection. Consider the determinable ``spin$_x$''. Friends of the determinable based account should identify that with a particular operator. And indeed a projection operator. This cannot but be the identity operator when restricted to the spin $\frac{1}{2}$ superselection sector---and this spells trouble for the account. We are not considering this in detail as it seems that determinable theorists have already explicitly identified (see e.g., Calosi and Wilson (2021: 3312)) the relevant determinable simply with the $\hat{S}_{x} = \begin{pmatrix} 0 & 1\\1 & 0\end{pmatrix}$---a choice that Fletcher and Taylor themselves agree is open for the determinable theorist. Whether this choice can be further articulated and defended by reconstructing Pauli matrices from orthomodular lattices is a question we leave for another occasion.}



\section{Conclusion}

We have argued in this paper that Aristotelian accounts of QI---that reject \textsc{Bivalence}---are metametaphysically dispensable. This means that, aside from problems they encounter on their own, \textsc{Meta-Level Accounts} that require dropping \textsc{Bivalence} are not metametaphysically necessary. While we have not (yet) dismissed the \textsc{Object-Level Accounts} that drop \textsc{Bivalence}, we have made the case that \textsc{Bivalent} $\mathcal{QL}$ \textsc{Object-Level Accounts} are metametaphysically sufficient to model the kind of indeterminacy we are concerned with. The \textsc{Bivalent} $\mathcal{QL}$ \textit{determinable-based} account of QI is an example of such an account, which should therefore be seen perhaps as the front-runner overall.

\section*{Acknowledgments} \small

For comments and discussions of previous drafts of this paper we would like to thank [\textsc{Redacted}].

\section*{References} \small

Bacciagaluppi, G. (2009) ``Is logic empirical?'', in Gabbay D., D. Lehmann, and K. Engesser (eds.) \textit{Handbook of Quantum Logic}, Elsevier, Amsterdam, 49--78. 

\bigskip

\noindent Barnes Williams, R. (2011). A Theory of Metaphysical Indeterminacy. In Bennett, K. \& Zimmerman, D. W. (eds.), Oxford Studies in Metaphysics volume 6, Oxford: Oxford University Press, 103-148.

\bigskip

\noindent Bell, J. \& M. Hallett (1982) ``Logic, quantum logic and empiricism'', in \textit{Philosophy of Science}, 49, 355--379.

\bigskip

\noindent Beltrametti, E.G. and G. Cassinelli (1981) \textit{The Logic of Quantum Mechanics}, Addison-Wesley, Reading (MA).
 
\bigskip

\noindent Bigaj, T. (2001) ``Three-valued logic, indeterminacy and quantum mechanics'', in \textit{Journal of Philosophical Logic}, 30, 97--119.

\bigskip

\noindent Birkhoff, G. \& J. von Neumann (1936) ``The logic of quantum mechanics'' in \textit{Annals of Mathematics} 37, 823--843.

\bigskip

\noindent Bokulich, A. (2014). Metaphysical Indeterminacy, Properties, and Quantum Theory. Res Philosophica, 91(3), 449-475.

\bigskip

\noindent Calosi, C., and Mariani, C. (2020). Quantum Relational Indeterminacy. \textit{Studies in History and Philosophy of Modern Physics}. At: \url{https://www.sciencedirect.com/science/article/pii/S1355219820300940}.

\bigskip

\noindent Calosi, C. and Mariani, C. (2021). Quantum Indeterminacy. \textit{Philosophy Compass} 16 (4): e12731.

\bigskip

\noindent Calosi, C. and Wilson, J. (2019). Quantum Metaphysical Indeterminacy. \textit{Philosophical Studies} 176 (10): 2599-2627.

\bigskip

\noindent Calosi, C. and Wilson, J. (2021). Quantum Indeterminacy and the Double-Slit Experiment. \textit{Philosophical Studies}, 178: 3291–3317.

\bigskip

\noindent Calosi, C. and Wilson, J. (2022) Metaphysical Indeterminacy in the Multiverse. In Allori, V. (Ed.) \textit{Quantum Mechanics and Fundamentality: Naturalizing Quantum Theory between Scientific Realism and Ontological Indeterminacy}, Springer.

\bigskip

\noindent Cameron, R. (2015). \textit{The Moving Spotlight}. Oxford: Oxford University Press.

\bigskip

\noindent Carnap, R. (1943) \textit{Formalization of Logic}, Harvard University Press.

\bigskip

\noindent Correia, F. and Rosenkrantz, S. (2018). \textit{Nothing to Come}. Berlin: Springer.

\bigskip

\noindent Corti, A. (2021). Yet again, quantum indeterminacy is not worldly indecision \textit{Synthese} 199:5623-5643.

\bigskip

\noindent Dalla Chiara, M., R. Giuntini, and R. Greechie (2004) \textit{Reasoning in Quantum Theory. Sharp and Unsharp Quantum Logics}, Kluwer, Dordrecht.

\bigskip

\noindent Darby, G. (2010). Quantum Mechanics and Metaphysical Indeterminacy. Australasian Journal of Philosophy, 88(2), 227-245. doi:\url{10.1080/00048400903097786}.

\bigskip

\noindent Darby, G. (2014) Vague Objects in Quantum Mechanics? In: Akiba, K (ed.) \textit{Vague Objects and Vague Identity}. Springer.

\bigskip

\noindent Darby, G., \& Pickup, M. (2019). Modelling Deep Indeterminacy. Synthese. doi:\url{10.1007/s11229-019-02158-0}.

\bigskip

\noindent Demopoulos, W. (1974) \textit{On the Possibility Structure of Physical Systems}, PhD dissertation, University of Western Ontario. 

\bigskip

\noindent Demopoulos, W. (1976) ``The Possibility Structure of Physical Systems'',  in W. Harper and C. A. Hooker (eds.) \textit{Foundations and philosophy of statistical theories in the physical
sciences}, Reidel, Dordrecht, 55--80.

\bigskip

\noindent Dummett, M. (1976) ``Is logic empirical?'' in \textit{Truth and Other Enigmas}, Duckworth, London, 269--289, 1978.

\bigskip

\noindent Dunn, J. M., \& Hardegree, G. (2001). \textit{Algebraic methods in philosophical logic}, Oxford University Press.

\bigskip

\noindent Fletcher, S. and Taylor, D. (2021a) Quantum indeterminacy and the eigenstate-eigenvalue link. \textit{Synthese} 199: 11181-11212.

\bigskip

\noindent French, S., \& Krause, D. (2006). 
Identity in Physics: A Historical, Philosophical, and Formal Analysis. Oxford: Oxford University Press. doi:\url{10.1093/0199278245.001.0001}.

\bigskip

\noindent Glick , D. (2017). Against quantum indeterminacy. Thought: A Journal of Philosophy, 6(3), 204-213.

\bigskip

 \noindent Grandjean, V. (2021) How is the Asymmetry between the Open Future and the Fixed Past to be characterized? \textit{Synthese}: 1863-1886.

\bigskip

\noindent Hellman, G. (1980) ``Quantum logic and meaning'', in \textit{Proceedings of the PSA}, 2, 493--511.

\bigskip

\noindent Horvat, S. and I. D. Toader (2023) ``Quantum logic and meaning'', \url{https://doi.org/10.48550/arXiv.2304.08450}

\bigskip

\noindent Humberstone, L. (2011) \textit{The Connectives}, MIT Press, Cambridge (MA).

\bigskip

\noindent Kochen, S. and E. P. Specker (1967) ``The Problem of Hidden Variables in Quantum Mechanics'',  in C. A. Hooker (ed.) \textit{The Logico-Algebraic Approach to Quantum Mechanics}, I, Dordrecht, Reidel, 1975, 293--328.

\bigskip

\noindent Landsman, K. (2017) \textit{Foundations of Quantum Theory. From Classical Concepts to Operator Algebras}, Springer

\bigskip

\noindent Lowe, E. J. (1994). Vague Identity and Quantum Indeterminacy. Analysis, 54(2), 110-114. doi:\url{10.1093/analys/54.2.110}.

\bigskip

\noindent Mariani, C. 2021. Emergent Quantum Indeterminacy. \textit{Ratio}. At: \url{https://onlinelibrary.wiley.com/doi/10.1111/rati.12305}.

\bigskip

\noindent Mariani, C., Michels, R., \& Torrengo, G. (2021) Plural metaphysical supervaluationism. \textit{Inquiry}, 67(6), 2005–2042. \url{https://www.tandfonline.com/doi/full/10.1080/0020174X.2021.1982404}.

\bigskip

\noindent Pavi\v{c}i\'{c}, M. (1992) ``Bibliography on quantum logics and related structures'', in \textit{International Journal of Theoretical Physics}, 31, 373--455.

\bigskip

\noindent Quine, W.V.O. (1981) ``What Price Bivalence?'', in \textit{Journal of Philosophy}, 78, 90--95. 

\bigskip

\noindent Rumfitt, I. (2015) \textit{The Boundary Stones of Thought. An Essay in the Philosophy of Logic}, Oxford University Press.

\bigskip

\noindent Rumfitt, I. (2018) ``Bivalence and Determinacy'', in M. Glanzberg (ed.) \textit{The Oxford Handbook of Truth}, Oxford University Press

\bigskip

 \noindent Schroeren, D. 2021. Quantum Metaphysical Indeterminacy and the Ontological Foundations of Orthodoxy. \textit{Studies in History and Philosophy of Science Part A} 90: 235-246.

\bigskip

\noindent Skow, B. (2010). Deep metaphysical indeterminacy. Philosophical Quarterly, 60(241), 851-858. doi:\url{10.1111/j.1467-9213.2010.672.x}.

\bigskip

\noindent Stairs, A. (2016). ``Could logic be empirical? The Putnam-Kripke debate'' in Chubb, J. \textit{et. al}., \textit{Logic and Algebraic Structures in Quantum Computing and Information}, Cambridge University Press, 23--41.

\bigskip

\noindent Toader, I. D. (2023). Distribution can be Dropped: Reply to Rumfitt. \textit{Analysis}, \url{https://doi.org/10.1093/analys/anae093}

\bigskip

\noindent Torza, A. (2020). Quantum metaphysical indeterminacy and worldly incompleteness. \textit{Synthese}, 197 (10): 4251-4264.

\bigskip

\noindent Torza, A. (2021). Quantum metametaphysics. \textit{Synthese} 199 (3-4):1-25.

\bigskip

\noindent Torza, A. (2022). Derivative metaphysical indeterminacy
and quantum physics. In Allori, V. (Ed.) \textit{Quantum Mechanics and Fundamentality: Naturalizing Quantum Theory between Scientific Realism and Ontological Indeterminacy}, Springer.

\bigskip 

\noindent Wilson, J. (2013). A Determinable-Based Account of Metaphysical Indeterminacy. \textit{Inquiry}, 56(4), 359–385.

\end{document}